\documentclass[aps,prl,superscriptaddress,twocolumn,amssymb,amsmath]{revtex4-1}
\usepackage{graphicx}% Include figure files
\usepackage{dcolumn}% Align table columns on decimal point
\usepackage{bm}% bold math
\usepackage{braket}
\usepackage[dvipsnames]{xcolor}
\newcommand{\g}{\ket{g}}
\newcommand{\e}{\ket{e}}
\newcommand{\f}{\ket{f}}

\newcommand{\ra}{\rightarrow}

\newcommand{\eqnl}[2]{\begin{equation}\label{#1}#2\end{equation}}
\newcommand{\eqn}[1]{\begin{equation}#1\end{equation}}

\usepackage[colorlinks,citecolor=blue]{hyperref} %references with links

\usepackage{mathptmx} %PRL font

\begin{document}

\preprint{APS/123-QED}
\title{Measuring effective temperatures of qubits using correlations}% Force line breaks with \\
%\thanks{Try following the APS guidelines}%

\author{Anatoly Kulikov}
\email{a.kulikov@uq.edu.au}
\affiliation{ARC Centre of Excellence for Engineered Quantum Systems, Queensland 4072, Australia}
\affiliation{School of Mathematics and Physics, University of Queensland, St Lucia, Queensland 4072, Australia}

\author{Rohit Navarathna}
\affiliation{ARC Centre of Excellence for Engineered Quantum Systems, Queensland 4072, Australia}
\affiliation{School of Mathematics and Physics, University of Queensland, St Lucia, Queensland 4072, Australia}

\author{Arkady Fedorov}
\affiliation{ARC Centre of Excellence for Engineered Quantum Systems, Queensland 4072, Australia}
\affiliation{School of Mathematics and Physics, University of Queensland, St Lucia, Queensland 4072, Australia}

\date{\today}% It is always \today, today,

\begin{abstract}
Initialization of a qubit in a pure state is a prerequisite for quantum computer operation. Qubits are commonly initialized by cooling to their ground states through passive thermalization or by using active reset protocols. To accurately quantify the initialization one requires a tool to measure the excited state population with sufficient accuracy given that the spurious excited state population may not exceed a fraction of a percent. In this Letter we propose a new technique of finding the excited state population of a qubit using correlations between two sequential measurements. We experimentally implement the proposed technique using a circuit QED platform and compare its performance with previously developed techniques. Unlike other techniques, our method does not require  high-fidelity readout and does not involve the excited levels of the system outside of the qubit subspace. We experimentally demonstrated measurement of the spurious qubit population with accuracy of up to $0.01\%$. This accuracy enabled us to perform ``temperature spectroscopy" of the qubit which helps to shed light on sources of the decoherence.
%\begin{description}
%\item[DOI]
%\end{description}
\end{abstract}

\pacs{Valid PACS appear here}% PACS, the Physics and Astronomy
\maketitle

%\tableofcontents

%{\textit{Introduction.}}---
Residual population of the excited state of superconducting qubits has been  routinely measured to be many orders of magnitude higher than the one predicted from the Maxwell-Boltzmann (M-B) distribution with a temperature of a dilution refrigerator. For the temperature of  $\lesssim$ 20 mK and for qubit frequencies $\sim 5$~GHz one might expect the population of the excited state $P_e  < 10^{-5}$, while the measured values are much larger and might even exceed one per cent~\cite{Corcoles2011,Siddiqi2012Heralded,DiCarlo2012Init,Devoret2013Reset,Toyli2016Resonance,Egger2018,Heinsoo2018}. This unexpected increase of the effective temperature of a qubit is one of the factors limiting the fidelity of operations in superconducting quantum processors and may be also an indication of extra decoherence channels for the qubit. Potential reasons for this spurious population may include hot out-of-equilibrium quasi-particles~\cite{Martinis2013QP,Devoret2018QP,Bespalov2016QP,Henriques2019} generated by stray radiation~\cite{Barends2011,Corcoles2011} or cosmic rays \cite{Bespalov2016QP,Henriques2019} and microwave noise~\cite{Yeh2017,Krinner2019} from the higher stages of a dilution refrigerator. 

In order to quantify the quality of the state initialization  and, more importantly, to identify and eliminate the sources of spurious excitation, one needs to resolve the changes in the excited state population of a qubit within fractions of a per cent. Using dispersive measurement with quantum limited amplification provides high signal-to-noise ratio (SNR) sufficient for a single-shot readout of the qubit state and enables direct counting of the excited state population by repeated measurement. 
Due to technical restrictions it is not always possible to use quantum limited amplifiers and, sometimes, it is not possible to reach the required measurement contrast even in the presence of quantum-limited amplification. An alternative method involves the third level of a system employed as a qubit: the amplitude of Rabi oscillations between the first and second excited states can be used as a measure of excited state population~\cite{Devoret2013Reset,Oliver2015Residual}. This method cannot be applied if the higher levels are not accessible due to large discrepancy of transition frequencies or selection rules~\cite{Liu2014}.

In this Letter we introduce a method allowing to measure the excited state ($\e$-state) population (or effective temperature $T_{\rm eff}$) of a qubit using correlations between two sequential measurements. Utilizing the quantum non-demolition (QND) nature of the measurement, we lift the requirements for high-fidelity single-shot readout or for manipulations involving higher levels of the system employed as a qubit to measure its effective temperature. 
The accuracy limit of our method is not limited by SNR and is only sensitive to qubit decoherence and gate errors in the second order. Because of that we achieve the highest reported precision of the excited state population measurement with accuracy of $0.01\%$ and study its dependence on the qubit transition frequency.
Although our experimental demonstration is carried out on the platform of circuit quantum electrodynamics (QED) and transmon qubits, the method is generic and is applicable to any system where the QND measurement can be realised.
%, including spin qubits, trapped ions and NV vacancies in diamonds (??)[links]\cite{Nakajima2019_QND_spin}.

Our experimental system consists of a tunable-frequency superconducting qubit, called a transmon, coupled to a 3D microwave cavity. The cavity is employed to both carry the microwave pulses to manipulate the qubit and to readout its state. The transmon has a weakly anharmonic multi-level structure, and its two lowest energy eigenstates are used as the logical states $\g$ and $\e$ of a qubit. We are also using the next energy eigenstate $\f$ to realize the qutrit protocol mentioned above \cite{Devoret2013Reset,Oliver2015Residual} for comparison. The system is tuned to the dispersive regime, where the qubit $\g$-$\e$ transition frequency is far from the cavity transition frequency, so the standard dispersive readout method can be employed~\cite{Blais2004}. For low readout powers the dispersive readout has highly quantum non-demolition nature with with negligible contribution to qubit excitation due to readout. To achieve a high SNR to be able to readout a transmon state in a single-shot regime we use a Josephson parametric amplifier (JPA) similar to one described in Ref.~\cite{Eichler2014}. 

We first present the idea of our method using a notion of an abstract ideal quantum two-level system with an instant and noiseless quantum non-demolition measurement. We can define the measurement apparatus to yield a real value $V_g$ for the qubit in the ground state and $V_e$ for the qubit in the excited state. By repeating the same experiment many times the average value of the measurement response is expressed as
\eqn{ \left<V\right> =g^{(0)}= P_g V_g + P_e V_e\equiv \tilde V_g,}
where  $P_e$ is spurious $\e$-state population, $P_g=1-P_e$ is the ground state population and $g^{(0)}$ is zero-th order correlation function. 
Knowledge of $\langle V\rangle$ can be in principle sufficient to determine the excited state population $P_e$ if the responses $V_{g/e}$ are known. Unfortunately, these responses are generally not known {\it a priori} and to determine $P_e$ one needs to make additional measurements such as some measurements involving the second excited level~\cite{Devoret2013Reset,Oliver2015Residual}. Instead of using the higher excited levels we propose to measure the first order correlation function $g^{(1)}(\tau)=\langle V(0)V(\tau)\rangle$. 

Assuming that our measurement is quantum non-demolition (QND) the second subsequent measurement will return a fully correlated result 
\eqn{g^{(1)}(0) = P_g V_g^2 + P_e V_e^2.\label{A}}
This value can be compared to 
\eqn{g^{(1)}(\infty) = \left(g^{(0)}\right)^2\leq g^{(1)}(0),\label{B}}
where we assumed that the measurements will be fully uncorrelated if separated by long times. It is also straightforward to see that the equality $g^{(1)}(0)=g^{(1)}(\infty)$ is realized only if the qubit is its ground state $P_g=1$ (or $P_e=1$). 
\begin{figure}[t]
	\centering
	\includegraphics[width=0.49\textwidth]{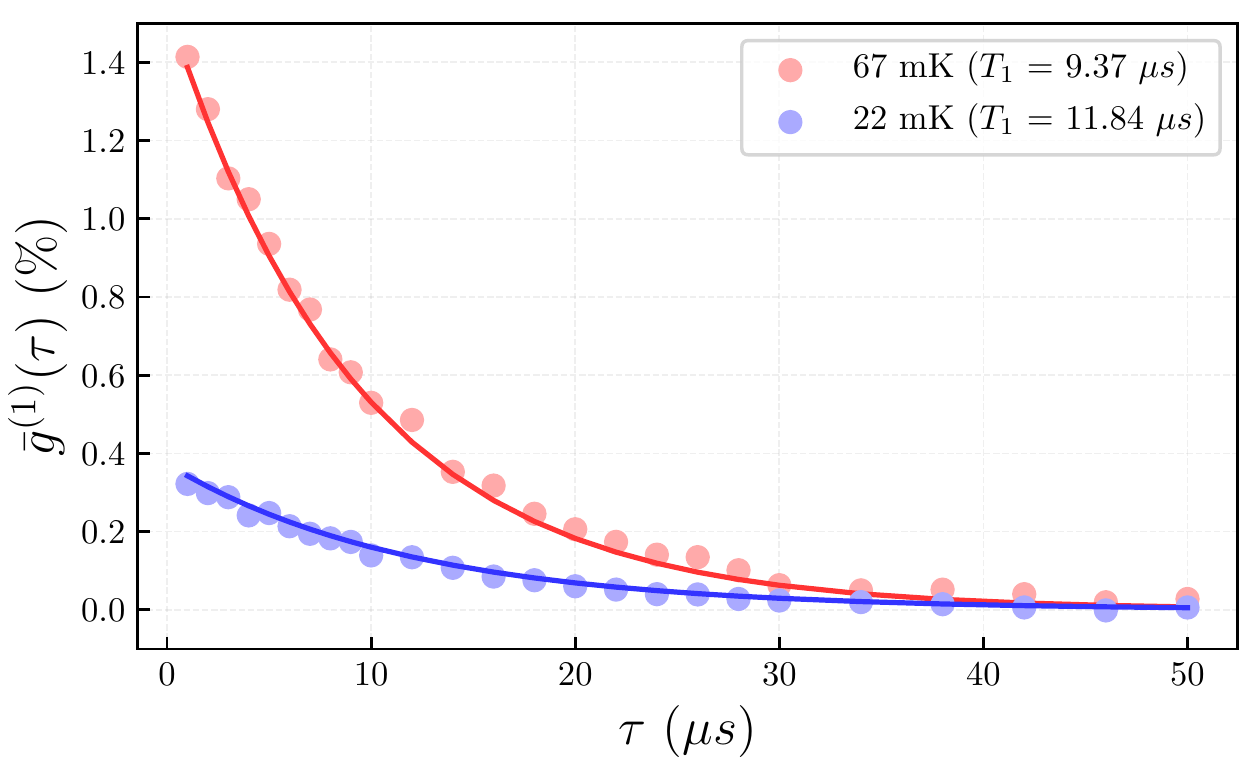}
	\caption{Decay of the normalized correlator between two sequential measurements separated by $\tau$. The solid lines represent exponential $T_1$-decay. Amplitude of the correlator at zero (or lowest attainable) delay allows one to reconstruct the $\e$-state population and hence the effective temperature of the qubit.
	}
	\label{fig:gaps}
\end{figure}

\begin{figure}[t]
	\centering
	\includegraphics[width=0.49\textwidth]{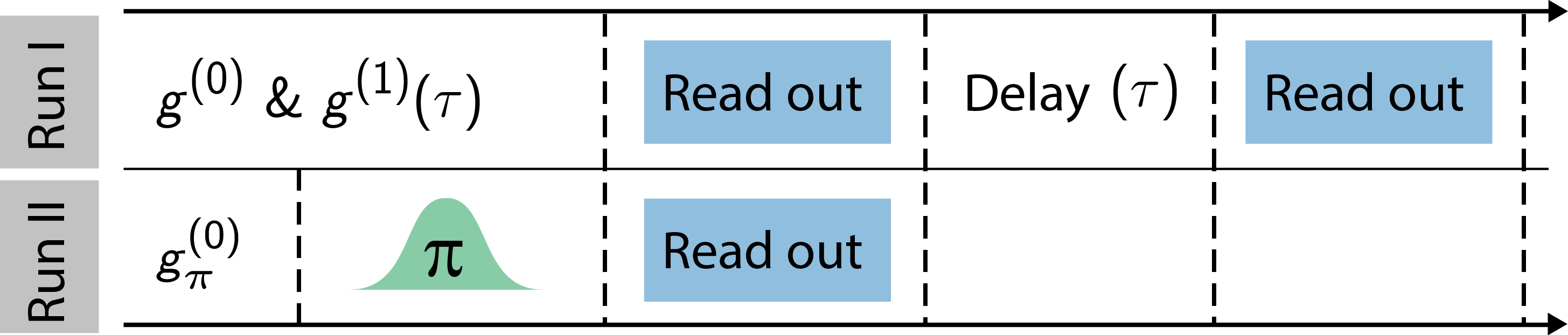}
	\caption{The experimental protocol. ``Run I" represents measurement of the correlation function $g^{(1)}(\tau)$ and $g^{(0)}$. ``Run II" is an additional calibration measurement required for correct scaling of $P_e$. The variable delay was used to measure the decay of $g^{(1)}(\tau)$. To determine $P_e$ only one measurement with $\tau=0$ is necessary.
	}
	\label{fig:protocol}
\end{figure}

Measurement of a typical decay of the correlation function is shown in Fig.~\ref{fig:gaps}; It follows an exponential curve with the relaxation time $T_1$ of the qubit. Observation of this decay is the manifestation of the spurious $\e$-state qubit population. However, to determine $P_e$ quantitatively we need to add a calibration measurement. For example, we can apply a $\pi$-pulse to swap the ground and excited state populations before taking a measurement (see Fig.~\ref{fig:protocol}) returning 
\eqn{g^{(0)}_\pi = P_e V_g + P_g V_e\equiv \tilde V_e.\label{C}}
Using simple calculations and an assumption of $P_e$ being small (see Supplemental material \cite{Supp}) one can obtain
\eqnl{eqn:p_e_base}{P_e \simeq \frac{g^{(1)}(0) - \left(g^{(0)}\right)^2}{\left(g^{(0)}+g^{(0)}_\pi - 2\sqrt{g^{(1)}(0)}\right)^2}.}

In circuit QED platform, we use the integrated heterodyne voltage transmitted through a resonator as an output of measurement apparatus. Heterodyne voltage is complex-valued, so we can use both quadratures as real-valued responses of the measurement apparatus. In practice, it is easier to work with a normalized real voltage ${\bar V} = \operatorname{Re} \left[ (V-\tilde V_g)/(\tilde V_e-\tilde V_g) \right]$ which is dimensionless and is defined to have the maximal distance between the ground and excited state responses. The zero-th order correlation functions of $\bar{V}$ are of a particularly simple form: $\bar{g}^{(0)}\equiv P_g \bar{V_g}+P_e \bar{V_e}=0$ and $\bar{g}^{(0)}_\pi\equiv P_e \bar{V_g}+P_g \bar{V_e}=1$. That allows us to write an exact simple expression for the spurious $\e$-state population (see Supplemental Material \cite{Supp}) as
\eqn{P_e = \frac{1}{2}-\frac{1}{2 \sqrt{1+4 \bar{g}^{(1)}(0)}}\simeq\bar{g}^{(1)}(0),}
 where $\bar{g}^{(1)}(0)\equiv\langle \bar{V}(0)\bar{V}(\tau)\rangle|_{\tau = 0}$ and the approximation holds when $P_e\ll1$. 
 
In reality, measurement of the correlation function returns $\bar{g}^{(1)}(\tau)=\langle \bar{V}(0)\bar{V}(\tau)\rangle +\langle \eta(0)\eta(\tau)\rangle$, where $\eta$ includes contributions of all noise sources such as noise of the amplification chain and the quantum noise. For a typical experimental setup the measurement noise is ``fast'' and  $\langle \eta(0)\eta(\tau)\rangle$ can be expressed as $\langle \eta\rangle^2=0$ for all relevant time scales. The noise contribution can be suppressed by acquiring sufficient statistics for all $\tau>0$. The noise contribution at $\tau=0$ can be, in principle, subtracted by performing additional calibration measurement of $\langle \eta^2\rangle$. In our experiments, we simply approximated $\bar{g}^{(1)}(0)$ by a correlator of the results of two sequential measurement in time (see Fig.~\ref{fig:protocol}). Systematic study of the standard deviation of $\bar{g}^{(1)}(0)$ shows the expected scaling with a number of averages $N$ up to $N=2^{16}$ confirming the absence of any measurable ``slow" noise contribution in our measurement setup (see below).

\begin{figure}[t]
	\centering
	\includegraphics[width=0.49\textwidth]{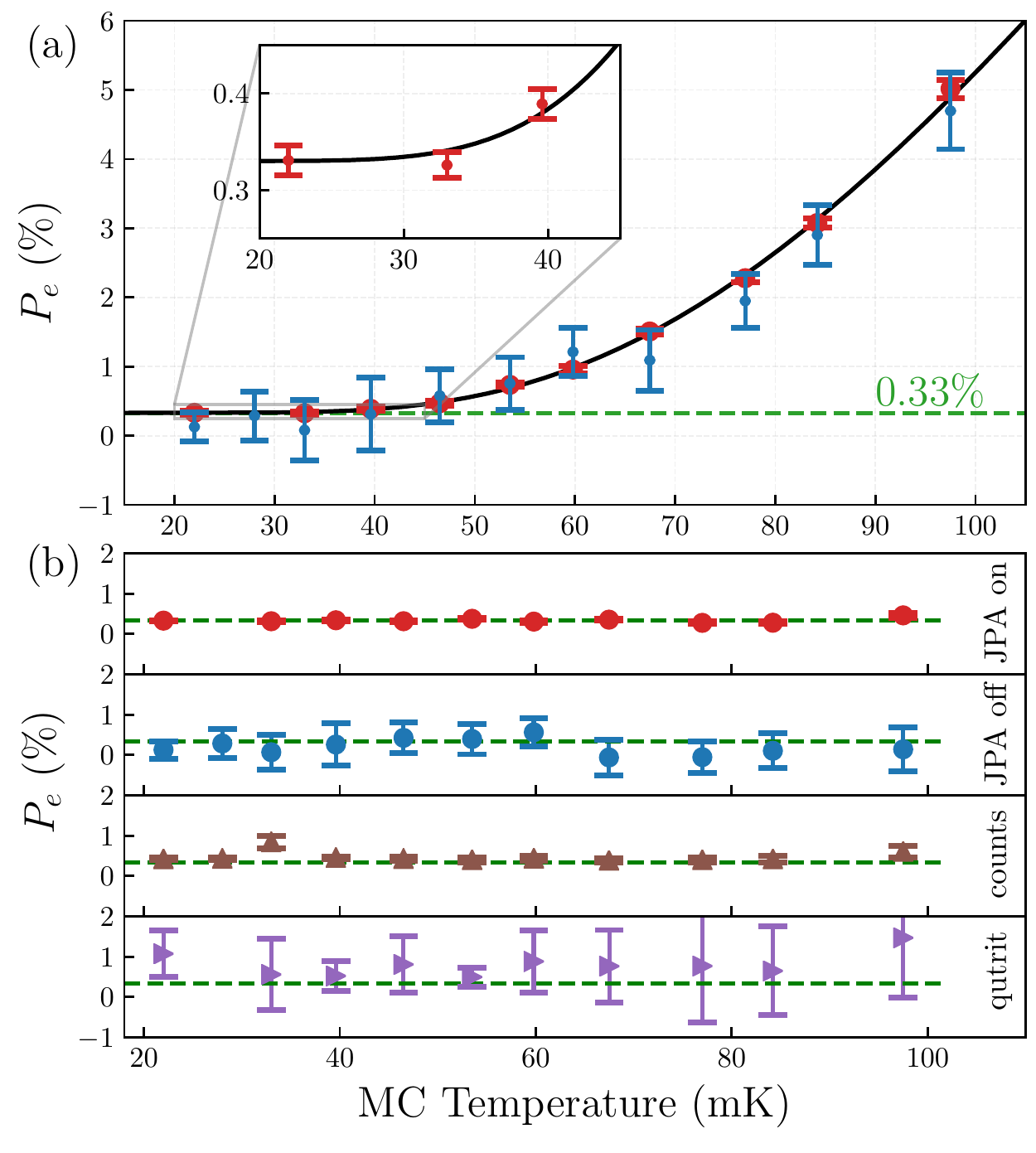}%{pic/temp_vs_pop_comparison_inset.pdf}%{pic/temp_vs_pop.pdf}
	\caption{(a) Measured $\e$-state population as a function of the mixing chamber sensor temperature. Red points are correlator measurements with a JPA. Data for each point corresponds to $2^{20}$ repetitions. The blue points are measured with JPA turned off. The black solid line corresponds to the M-B distribution offset by $0.33\%$ as indicated by the dashed green line. The error-bars cover two standard deviations in measurement ($95\%$ confidence). (b) Deviation of the data from the M-B distribution for different methods (see text for more details).}
	\label{fig:T_vs_T}
\end{figure}

{\it Results.---}We have performed a study of residual excited state population of a Transmon qubit vs the temperature of the mixing chamber (MC) plate of a dilution refrigerator shown in Fig.~\ref{fig:T_vs_T}. For each temperature point after stabilizing the MC sensor temperature we have waited ample time ($>$~1 hour) for the qubit and its environment to thermalize and performed measurement of the qubit $\e$-state population using four different methods for each MC temperature point. First, we used our method in the presence of a quantum-limited amplifier (JPA), which gives us a fairly high SNR of $\sim 6$ and allows determining the residual $\e$-state population with the precision of $.01\%$ in 15 minutes, which is the highest precision reported \cite{Oliver2015Residual,Heinsoo2018}. Interestingly, the standard deviation of our method was smaller than the direct counting of excitations using the same data.

In the second measurement we used our method without the JPA. It resulted in a SNR of ~0.9 which is not sufficient for a single-shot measurement. The results were in agreement with the precise measurements, thus demonstrating the ability of our method to work in the conditions of low SNR (Fig.~\ref{fig:T_vs_T}a). We have also used conventional methods to determine the $\e$-state population using the second excited state of the Transmon and the direct count of single shots making use of JPA~\cite{Eichler2014}. All methods' results are in agreement within the error bars but show different statistical and systematic errors (see Fig.~\ref{fig:T_vs_T}b and below for more comments).

The residual $\e$-state population of our qubit as function of the temperature of MC plate coincides within the error bars ($< 0.01\%$ uncertainty) with the M-B curve shifted by a ``zero-temperature excitation" offset (the curve is indicated on the plot with a solid black line). Note that both the offset value and the M-B distribution have no free parameters: the offset is given by the measurement at the lowest attainable temperature and the qubit transition energy was obtained independently using spectroscopy and Ramsey-type measurement. Our results are somewhat different from the conclusion of Ref.~\cite{Oliver2015Residual} where spurious excitation followed the M-B distribution without an offset, but saturated at the temperature of 35~mK.

The presence of this offset may be explained by a model of a qubit being coupled to two separate thermal baths. One of the baths is strongly coupled to the qubit and thermalised with the MC plate of a dilution refrigerator, while the second bath is weakly coupled but has a much higher temperature independent of the MC temperature. We determined the rate of excitation and relaxation events from this second, non-equilibrium source, to not exceed 670 Hz, corresponding to a time constant of 1.5 ms which is consistent with 'hot' out-of-equilibrium quasiparticles as a possible origin for the qubit excitation  \cite{Devoret2018QP,Aumentado2004}. 

\begin{figure}[t]
	\centering
	\includegraphics[width=0.49\textwidth]{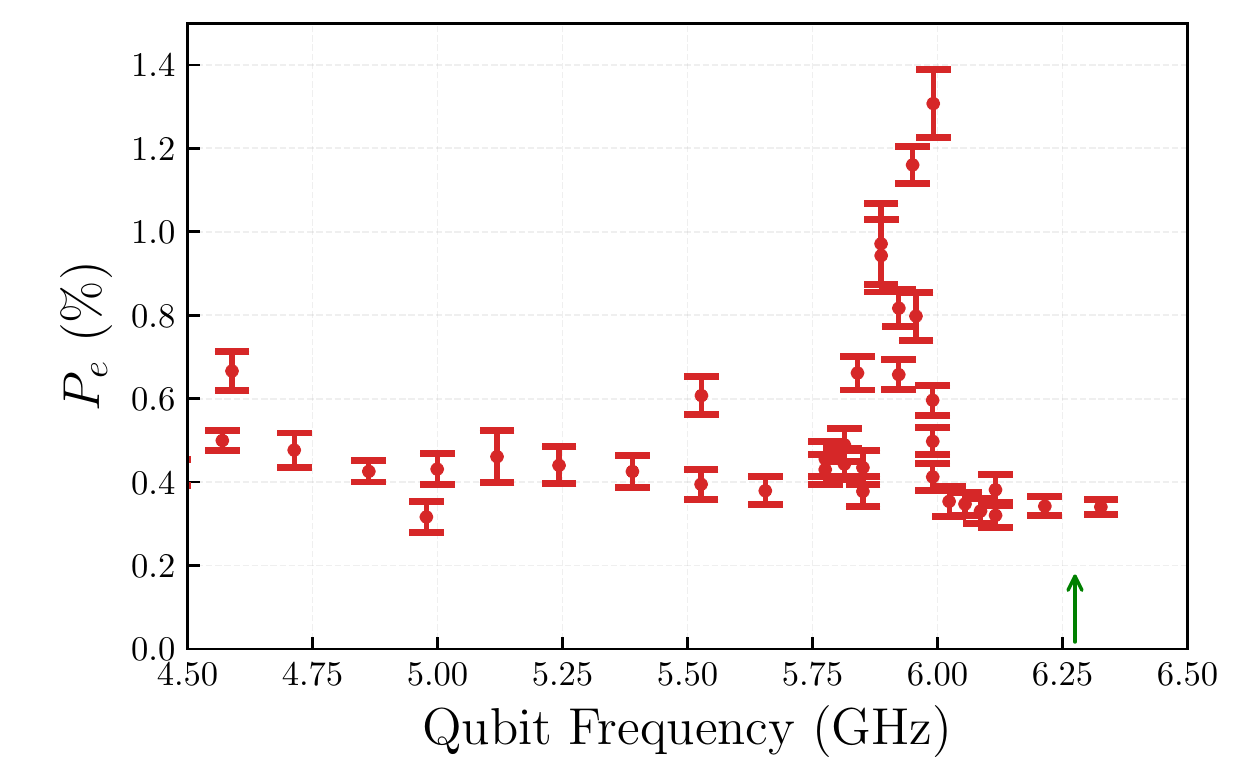}
	\caption{Excited state population vs qubit frequency representing a ``noise spectrum" as seen by the qubit. The green arrow indicates the qubit frequency used for the rest of the experiments.}
	\label{fig:pop_vs_freq1}
\end{figure}

To acquire more information on the origin of the qubit excitation we used our method to perform ``temperature spectroscopy" by measuring the $\e$-state population as a function of the qubit frequency. Fig.~\ref{fig:pop_vs_freq1} shows that the $\e$-state population peaks around 6~GHz and can change abruptly with even small changes in the qubit frequency. This behaviour is inconsistent with the excitation by quasiparticles whose matrix element is a smooth function of qubit frequency~\cite{Catelani2011}. Instead, this behaviour is characteristic to coupling to two-level systems (TLS) which are believed to be the dominant source of the qubit relaxation and exhibit a strong non-monotonic dependence of relaxation times of superconducting qubits on their frequencies~\cite{Klimov2018}.

{\it Precision and errors.---} 
Direct counting of $\e$-state population with single-shot readouts provides the most direct method of $\e$-state population measurement without use of any control pulses and was very instructive for a reliable verification of our method. Unfortunately, direct counting is only possible for a readout with sufficiently large SNR. With lower SNR the absolute error due to state misinterpretation rises exponentially thus limiting the practicality of this method for temperature measurement, especially for very small spurious populations.

The largest systematic error source of our method comes from the finite time of the measurement, which leads to a partial decay of the correlations following the standard T$_1$ decay curve (see Fig.~\ref{fig:gaps}). While this error can be considerable,  a separate measurement of T$_1$ can be used to correct for this error. Most importantly, this error is relative, as it only decreases the measured $P_e$ by a factor of $e^{-T_{\rm meas}/T_1}$, where $T_{\rm meas}$  is the measurement time. Therefore, this error does not set a lower limit on the measurable spurious population unlike the error of the finite SNR for the direct counting. 

A similar effect is due to $\pi$-pulse errors. As this error only affects $\tilde  V_e$ which is measured independently from $g^{(1)}$, it only contributes as a relative error and does not affect statistical distribution for $P_e$. Moreover, if an infidelity of the $\pi$-pulse is small this error contributes to $P_e$ only in the second order. 

Similarly to the direct counting method, measurement of $g^{(1)}$ does not involve any control pulses and is generally performed when the qubit is in equilibrium with environment. Therefore, the only possible systematic absolute error of our method arises from the excitation of the qubit due to dispersive readout which can be virtually arbitrarily suppressed by larger qubit detunings and/or lower readout powers. 

\begin{figure}[t]
	\centering
	\includegraphics[width=0.49\textwidth]{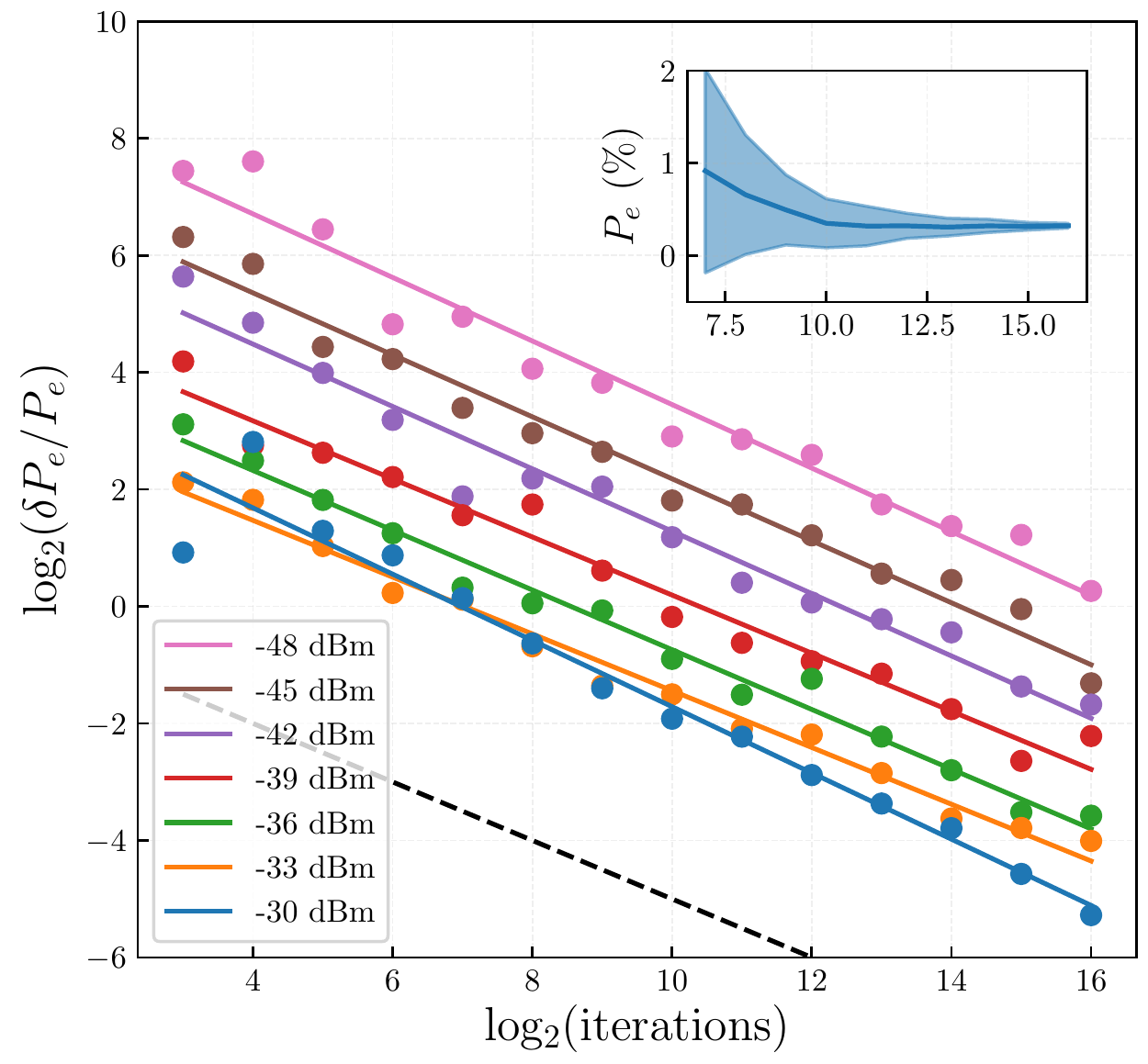}
	\caption{Relative precision of $P_e$ measurements and their linear fits. The precision scales as expected for uncorrelated noise (indicated by the black dashed line). Inset: Population (solid line) and standard deviation (fill) for a measurement power of $-30$ dBm.}
	\label{fig:snr_sweep}
\end{figure}

The only statistical (not systematic) error of our method is due to measurement noise which, in turn, can be reduced by increase in averaging time. Fig.~\ref{fig:snr_sweep} shows a standard deviation of measured $\e$-state population as a function of number of measurements and different readout powers. The error scales as $N^{-1/2}$, where $N$ is the number of iterations, over the complete range deviating from this expected dependence only for the largest power of -30~dBm, most probably, due to loss of quantum non-demolition behaviour of the readout.

It is interesting to note that the qutrit method demonstrated the worst accuracy which may be attributed to extra decoherence due to $\f$-level and to the direct excitation of $\e$-state when applying e-f drive. While certain optimal control techniques, such as DRAG pulses \cite{Motzoi2009} for e-f transition, could be employed to mitigate this problem, impossibility to entirely isolate spurious $\e$-state excitation by the method itself poses an extra limitation on its absolute precision.

{\it Discussions.---} In summary, we have proposed and experimentally realized a method of measuring the effective temperature of qubits using correlations between consecutive measurements. Our method does not require usage of higher excited levels, is less susceptible to errors in control pulses and allows for virtually unlimited suppression of absolute errors even without high SNR required for the high-fidelity single-shot measurement. Our method can be used on any platform; We experimentally show it to have the highest reported precision for superconducting circuits. The accuracy of our method enables ``temperature spectroscopy" giving spurious population of $\e$-state of the qubit as function of qubit transition frequency which can shed light on the sources of decoherence. 

\begin{acknowledgements}
We thank G. Milburn and B. Huard %for critical reading of our manuscript 
for fruitful discussions. We thank Andreas Wallraff, Markus Oppliger, Anton Poto\v{c}nik, and Mintu Mondal for fabricating JPA used in the measurements. The authors were supported by the Australian Research Council Centre of Excellence for Engineered Quantum Systems (EQUS, CE170100009). 
\end{acknowledgements}


%\nocite{*}

\bibliographystyle{apsrev4-1}
%\bibliography{draft_v_2.3}% Produces the bibliography via BibTeX.
\bibliography{RefDB/SQDRefDB,RefDB/extra_refs}

\providecommand{\noopsort}[1]{}\providecommand{\singleletter}[1]{#1}%
\begin{thebibliography}{25}%
\makeatletter
\providecommand \@ifxundefined [1]{%
 \@ifx{#1\undefined}
}%
\providecommand \@ifnum [1]{%
 \ifnum #1\expandafter \@firstoftwo
 \else \expandafter \@secondoftwo
 \fi
}%
\providecommand \@ifx [1]{%
 \ifx #1\expandafter \@firstoftwo
 \else \expandafter \@secondoftwo
 \fi
}%
\providecommand \natexlab [1]{#1}%
\providecommand \enquote  [1]{``#1''}%
\providecommand \bibnamefont  [1]{#1}%
\providecommand \bibfnamefont [1]{#1}%
\providecommand \citenamefont [1]{#1}%
\providecommand \href@noop [0]{\@secondoftwo}%
\providecommand \href [0]{\begingroup \@sanitize@url \@href}%
\providecommand \@href[1]{\@@startlink{#1}\@@href}%
\providecommand \@@href[1]{\endgroup#1\@@endlink}%
\providecommand \@sanitize@url [0]{\catcode `\\12\catcode `\$12\catcode
  `\&12\catcode `\#12\catcode `\^12\catcode `\_12\catcode `\%12\relax}%
\providecommand \@@startlink[1]{}%
\providecommand \@@endlink[0]{}%
\providecommand \url  [0]{\begingroup\@sanitize@url \@url }%
\providecommand \@url [1]{\endgroup\@href {#1}{\urlprefix }}%
\providecommand \urlprefix  [0]{URL }%
\providecommand \Eprint [0]{\href }%
\providecommand \doibase [0]{http://dx.doi.org/}%
\providecommand \selectlanguage [0]{\@gobble}%
\providecommand \bibinfo  [0]{\@secondoftwo}%
\providecommand \bibfield  [0]{\@secondoftwo}%
\providecommand \translation [1]{[#1]}%
\providecommand \BibitemOpen [0]{}%
\providecommand \bibitemStop [0]{}%
\providecommand \bibitemNoStop [0]{.\EOS\space}%
\providecommand \EOS [0]{\spacefactor3000\relax}%
\providecommand \BibitemShut  [1]{\csname bibitem#1\endcsname}%
\let\auto@bib@innerbib\@empty
%</preamble>
\bibitem [{\citenamefont {C\'{o}rcoles}\ \emph {et~al.}(2011)\citenamefont
  {C\'{o}rcoles}, \citenamefont {Chow}, \citenamefont {Gambetta}, \citenamefont
  {Rigetti}, \citenamefont {Rozen}, \citenamefont {Keefe}, \citenamefont
  {Rothwell}, \citenamefont {Ketchen},\ and\ \citenamefont
  {Steffen}}]{Corcoles2011}%
  \BibitemOpen
  \bibfield  {author} {\bibinfo {author} {\bibfnamefont {A.~D.}\ \bibnamefont
  {C\'{o}rcoles}}, \bibinfo {author} {\bibfnamefont {J.~M.}\ \bibnamefont
  {Chow}}, \bibinfo {author} {\bibfnamefont {J.~M.}\ \bibnamefont {Gambetta}},
  \bibinfo {author} {\bibfnamefont {C.}~\bibnamefont {Rigetti}}, \bibinfo
  {author} {\bibfnamefont {J.~R.}\ \bibnamefont {Rozen}}, \bibinfo {author}
  {\bibfnamefont {G.~A.}\ \bibnamefont {Keefe}}, \bibinfo {author}
  {\bibfnamefont {M.~B.}\ \bibnamefont {Rothwell}}, \bibinfo {author}
  {\bibfnamefont {M.~B.}\ \bibnamefont {Ketchen}}, \ and\ \bibinfo {author}
  {\bibfnamefont {M.}~\bibnamefont {Steffen}},\ }\href {\doibase
  10.1063/1.3658630} {\bibfield  {journal} {\bibinfo  {journal} {Appl. Phys.
  Lett.}\ }\textbf {\bibinfo {volume} {99}},\ \bibinfo {eid} {181906} (\bibinfo
  {year} {2011})}\BibitemShut {NoStop}%
\bibitem [{\citenamefont {Johnson}\ \emph {et~al.}(2012)\citenamefont
  {Johnson}, \citenamefont {Macklin}, \citenamefont {Slichter}, \citenamefont
  {Vijay}, \citenamefont {Weingarten}, \citenamefont {Clarke},\ and\
  \citenamefont {Siddiqi}}]{Siddiqi2012Heralded}%
  \BibitemOpen
  \bibfield  {author} {\bibinfo {author} {\bibfnamefont {J.~E.}\ \bibnamefont
  {Johnson}}, \bibinfo {author} {\bibfnamefont {C.}~\bibnamefont {Macklin}},
  \bibinfo {author} {\bibfnamefont {D.~H.}\ \bibnamefont {Slichter}}, \bibinfo
  {author} {\bibfnamefont {R.}~\bibnamefont {Vijay}}, \bibinfo {author}
  {\bibfnamefont {E.~B.}\ \bibnamefont {Weingarten}}, \bibinfo {author}
  {\bibfnamefont {J.}~\bibnamefont {Clarke}}, \ and\ \bibinfo {author}
  {\bibfnamefont {I.}~\bibnamefont {Siddiqi}},\ }\href {\doibase
  10.1103/PhysRevLett.109.050506} {\bibfield  {journal} {\bibinfo  {journal}
  {Phys. Rev. Lett.}\ }\textbf {\bibinfo {volume} {109}},\ \bibinfo {pages}
  {050506} (\bibinfo {year} {2012})}\BibitemShut {NoStop}%
\bibitem [{\citenamefont {Rist\`e}\ \emph {et~al.}(2012)\citenamefont
  {Rist\`e}, \citenamefont {van Leeuwen}, \citenamefont {Ku}, \citenamefont
  {Lehnert},\ and\ \citenamefont {DiCarlo}}]{DiCarlo2012Init}%
  \BibitemOpen
  \bibfield  {author} {\bibinfo {author} {\bibfnamefont {D.}~\bibnamefont
  {Rist\`e}}, \bibinfo {author} {\bibfnamefont {J.~G.}\ \bibnamefont {van
  Leeuwen}}, \bibinfo {author} {\bibfnamefont {H.-S.}\ \bibnamefont {Ku}},
  \bibinfo {author} {\bibfnamefont {K.~W.}\ \bibnamefont {Lehnert}}, \ and\
  \bibinfo {author} {\bibfnamefont {L.}~\bibnamefont {DiCarlo}},\ }\href
  {\doibase 10.1103/PhysRevLett.109.050507} {\bibfield  {journal} {\bibinfo
  {journal} {Phys. Rev. Lett.}\ }\textbf {\bibinfo {volume} {109}},\ \bibinfo
  {pages} {050507} (\bibinfo {year} {2012})}\BibitemShut {NoStop}%
\bibitem [{\citenamefont {Geerlings}\ \emph {et~al.}(2013)\citenamefont
  {Geerlings}, \citenamefont {Leghtas}, \citenamefont {Pop}, \citenamefont
  {Shankar}, \citenamefont {Frunzio}, \citenamefont {Schoelkopf}, \citenamefont
  {Mirrahimi},\ and\ \citenamefont {Devoret}}]{Devoret2013Reset}%
  \BibitemOpen
  \bibfield  {author} {\bibinfo {author} {\bibfnamefont {K.}~\bibnamefont
  {Geerlings}}, \bibinfo {author} {\bibfnamefont {Z.}~\bibnamefont {Leghtas}},
  \bibinfo {author} {\bibfnamefont {I.~M.}\ \bibnamefont {Pop}}, \bibinfo
  {author} {\bibfnamefont {S.}~\bibnamefont {Shankar}}, \bibinfo {author}
  {\bibfnamefont {L.}~\bibnamefont {Frunzio}}, \bibinfo {author} {\bibfnamefont
  {R.~J.}\ \bibnamefont {Schoelkopf}}, \bibinfo {author} {\bibfnamefont
  {M.}~\bibnamefont {Mirrahimi}}, \ and\ \bibinfo {author} {\bibfnamefont
  {M.~H.}\ \bibnamefont {Devoret}},\ }\href {\doibase
  10.1103/PhysRevLett.110.120501} {\bibfield  {journal} {\bibinfo  {journal}
  {Phys. Rev. Lett.}\ }\textbf {\bibinfo {volume} {110}},\ \bibinfo {pages}
  {120501} (\bibinfo {year} {2013})}\BibitemShut {NoStop}%
\bibitem [{\citenamefont {Toyli}\ \emph {et~al.}(2016)\citenamefont {Toyli},
  \citenamefont {Eddins}, \citenamefont {Boutin}, \citenamefont {Puri},
  \citenamefont {Hover}, \citenamefont {Bolkhovsky}, \citenamefont {Oliver},
  \citenamefont {Blais},\ and\ \citenamefont {Siddiqi}}]{Toyli2016Resonance}%
  \BibitemOpen
  \bibfield  {author} {\bibinfo {author} {\bibfnamefont {D.~M.}\ \bibnamefont
  {Toyli}}, \bibinfo {author} {\bibfnamefont {A.~W.}\ \bibnamefont {Eddins}},
  \bibinfo {author} {\bibfnamefont {S.}~\bibnamefont {Boutin}}, \bibinfo
  {author} {\bibfnamefont {S.}~\bibnamefont {Puri}}, \bibinfo {author}
  {\bibfnamefont {D.}~\bibnamefont {Hover}}, \bibinfo {author} {\bibfnamefont
  {V.}~\bibnamefont {Bolkhovsky}}, \bibinfo {author} {\bibfnamefont {W.~D.}\
  \bibnamefont {Oliver}}, \bibinfo {author} {\bibfnamefont {A.}~\bibnamefont
  {Blais}}, \ and\ \bibinfo {author} {\bibfnamefont {I.}~\bibnamefont
  {Siddiqi}},\ }\href {\doibase 10.1103/PhysRevX.6.031004} {\bibfield
  {journal} {\bibinfo  {journal} {Phys. Rev. X}\ }\textbf {\bibinfo {volume}
  {6}},\ \bibinfo {pages} {031004} (\bibinfo {year} {2016})}\BibitemShut
  {NoStop}%
\bibitem [{\citenamefont {Egger}\ \emph {et~al.}(2018)\citenamefont {Egger},
  \citenamefont {Werninghaus}, \citenamefont {Ganzhorn}, \citenamefont {Salis},
  \citenamefont {Fuhrer}, \citenamefont {M\"uller},\ and\ \citenamefont
  {Filipp}}]{Egger2018}%
  \BibitemOpen
  \bibfield  {author} {\bibinfo {author} {\bibfnamefont {D.}~\bibnamefont
  {Egger}}, \bibinfo {author} {\bibfnamefont {M.}~\bibnamefont {Werninghaus}},
  \bibinfo {author} {\bibfnamefont {M.}~\bibnamefont {Ganzhorn}}, \bibinfo
  {author} {\bibfnamefont {G.}~\bibnamefont {Salis}}, \bibinfo {author}
  {\bibfnamefont {A.}~\bibnamefont {Fuhrer}}, \bibinfo {author} {\bibfnamefont
  {P.}~\bibnamefont {M\"uller}}, \ and\ \bibinfo {author} {\bibfnamefont
  {S.}~\bibnamefont {Filipp}},\ }\href {\doibase
  10.1103/PhysRevApplied.10.044030} {\bibfield  {journal} {\bibinfo  {journal}
  {Phys. Rev. Applied}\ }\textbf {\bibinfo {volume} {10}},\ \bibinfo {pages}
  {044030} (\bibinfo {year} {2018})}\BibitemShut {NoStop}%
\bibitem [{\citenamefont {Heinsoo}\ \emph {et~al.}(2018)\citenamefont
  {Heinsoo}, \citenamefont {Andersen}, \citenamefont {Remm}, \citenamefont
  {Krinner}, \citenamefont {Walter}, \citenamefont {Salath\'e}, \citenamefont
  {Gasparinetti}, \citenamefont {Besse}, \citenamefont
  {Poto\ifmmode~\check{c}\else \v{c}\fi{}nik}, \citenamefont {Wallraff},\ and\
  \citenamefont {Eichler}}]{Heinsoo2018}%
  \BibitemOpen
  \bibfield  {author} {\bibinfo {author} {\bibfnamefont {J.}~\bibnamefont
  {Heinsoo}}, \bibinfo {author} {\bibfnamefont {C.~K.}\ \bibnamefont
  {Andersen}}, \bibinfo {author} {\bibfnamefont {A.}~\bibnamefont {Remm}},
  \bibinfo {author} {\bibfnamefont {S.}~\bibnamefont {Krinner}}, \bibinfo
  {author} {\bibfnamefont {T.}~\bibnamefont {Walter}}, \bibinfo {author}
  {\bibfnamefont {Y.}~\bibnamefont {Salath\'e}}, \bibinfo {author}
  {\bibfnamefont {S.}~\bibnamefont {Gasparinetti}}, \bibinfo {author}
  {\bibfnamefont {J.-C.}\ \bibnamefont {Besse}}, \bibinfo {author}
  {\bibfnamefont {A.}~\bibnamefont {Poto\ifmmode~\check{c}\else
  \v{c}\fi{}nik}}, \bibinfo {author} {\bibfnamefont {A.}~\bibnamefont
  {Wallraff}}, \ and\ \bibinfo {author} {\bibfnamefont {C.}~\bibnamefont
  {Eichler}},\ }\href {\doibase 10.1103/PhysRevApplied.10.034040} {\bibfield
  {journal} {\bibinfo  {journal} {Phys. Rev. Applied}\ }\textbf {\bibinfo
  {volume} {10}},\ \bibinfo {pages} {034040} (\bibinfo {year}
  {2018})}\BibitemShut {NoStop}%
\bibitem [{\citenamefont {Wenner}\ \emph {et~al.}(2013)\citenamefont {Wenner},
  \citenamefont {Yin}, \citenamefont {Lucero}, \citenamefont {Barends},
  \citenamefont {Chen}, \citenamefont {Chiaro}, \citenamefont {Kelly},
  \citenamefont {Lenander}, \citenamefont {Mariantoni}, \citenamefont
  {Megrant}, \citenamefont {Neill}, \citenamefont {O'Malley}, \citenamefont
  {Sank}, \citenamefont {Vainsencher}, \citenamefont {Wang}, \citenamefont
  {White}, \citenamefont {Cleland},\ and\ \citenamefont
  {Martinis}}]{Martinis2013QP}%
  \BibitemOpen
  \bibfield  {author} {\bibinfo {author} {\bibfnamefont {J.}~\bibnamefont
  {Wenner}}, \bibinfo {author} {\bibfnamefont {Y.}~\bibnamefont {Yin}},
  \bibinfo {author} {\bibfnamefont {E.}~\bibnamefont {Lucero}}, \bibinfo
  {author} {\bibfnamefont {R.}~\bibnamefont {Barends}}, \bibinfo {author}
  {\bibfnamefont {Y.}~\bibnamefont {Chen}}, \bibinfo {author} {\bibfnamefont
  {B.}~\bibnamefont {Chiaro}}, \bibinfo {author} {\bibfnamefont
  {J.}~\bibnamefont {Kelly}}, \bibinfo {author} {\bibfnamefont
  {M.}~\bibnamefont {Lenander}}, \bibinfo {author} {\bibfnamefont
  {M.}~\bibnamefont {Mariantoni}}, \bibinfo {author} {\bibfnamefont
  {A.}~\bibnamefont {Megrant}}, \bibinfo {author} {\bibfnamefont
  {C.}~\bibnamefont {Neill}}, \bibinfo {author} {\bibfnamefont {P.~J.~J.}\
  \bibnamefont {O'Malley}}, \bibinfo {author} {\bibfnamefont {D.}~\bibnamefont
  {Sank}}, \bibinfo {author} {\bibfnamefont {A.}~\bibnamefont {Vainsencher}},
  \bibinfo {author} {\bibfnamefont {H.}~\bibnamefont {Wang}}, \bibinfo {author}
  {\bibfnamefont {T.~C.}\ \bibnamefont {White}}, \bibinfo {author}
  {\bibfnamefont {A.~N.}\ \bibnamefont {Cleland}}, \ and\ \bibinfo {author}
  {\bibfnamefont {J.~M.}\ \bibnamefont {Martinis}},\ }\href {\doibase
  10.1103/PhysRevLett.110.150502} {\bibfield  {journal} {\bibinfo  {journal}
  {Phys. Rev. Lett.}\ }\textbf {\bibinfo {volume} {110}},\ \bibinfo {pages}
  {150502} (\bibinfo {year} {2013})}\BibitemShut {NoStop}%
\bibitem [{\citenamefont {Serniak}\ \emph {et~al.}(2018)\citenamefont
  {Serniak}, \citenamefont {Hays}, \citenamefont {de~Lange}, \citenamefont
  {Diamond}, \citenamefont {Shankar}, \citenamefont {Burkhart}, \citenamefont
  {Frunzio}, \citenamefont {Houzet},\ and\ \citenamefont
  {Devoret}}]{Devoret2018QP}%
  \BibitemOpen
  \bibfield  {author} {\bibinfo {author} {\bibfnamefont {K.}~\bibnamefont
  {Serniak}}, \bibinfo {author} {\bibfnamefont {M.}~\bibnamefont {Hays}},
  \bibinfo {author} {\bibfnamefont {G.}~\bibnamefont {de~Lange}}, \bibinfo
  {author} {\bibfnamefont {S.}~\bibnamefont {Diamond}}, \bibinfo {author}
  {\bibfnamefont {S.}~\bibnamefont {Shankar}}, \bibinfo {author} {\bibfnamefont
  {L.~D.}\ \bibnamefont {Burkhart}}, \bibinfo {author} {\bibfnamefont
  {L.}~\bibnamefont {Frunzio}}, \bibinfo {author} {\bibfnamefont
  {M.}~\bibnamefont {Houzet}}, \ and\ \bibinfo {author} {\bibfnamefont {M.~H.}\
  \bibnamefont {Devoret}},\ }\href {\doibase 10.1103/PhysRevLett.121.157701}
  {\bibfield  {journal} {\bibinfo  {journal} {Phys. Rev. Lett.}\ }\textbf
  {\bibinfo {volume} {121}},\ \bibinfo {pages} {157701} (\bibinfo {year}
  {2018})}\BibitemShut {NoStop}%
\bibitem [{\citenamefont {Bespalov}\ \emph {et~al.}(2016)\citenamefont
  {Bespalov}, \citenamefont {Houzet}, \citenamefont {Meyer},\ and\
  \citenamefont {Nazarov}}]{Bespalov2016QP}%
  \BibitemOpen
  \bibfield  {author} {\bibinfo {author} {\bibfnamefont {A.}~\bibnamefont
  {Bespalov}}, \bibinfo {author} {\bibfnamefont {M.}~\bibnamefont {Houzet}},
  \bibinfo {author} {\bibfnamefont {J.~S.}\ \bibnamefont {Meyer}}, \ and\
  \bibinfo {author} {\bibfnamefont {Y.~V.}\ \bibnamefont {Nazarov}},\ }\href
  {\doibase 10.1103/PhysRevLett.117.117002} {\bibfield  {journal} {\bibinfo
  {journal} {Phys. Rev. Lett.}\ }\textbf {\bibinfo {volume} {117}},\ \bibinfo
  {pages} {117002} (\bibinfo {year} {2016})}\BibitemShut {NoStop}%
\bibitem [{\citenamefont {Henriques}\ \emph {et~al.}(2019)\citenamefont
  {Henriques}, \citenamefont {Valenti}, \citenamefont {Charpentier},
  \citenamefont {Lagoin}, \citenamefont {Gouriou}, \citenamefont
  {Mart{\'\i}nez}, \citenamefont {Cardani}, \citenamefont {Gr{\"u}nhaupt},
  \citenamefont {Gusenkova}, \citenamefont {Ferrero} \emph
  {et~al.}}]{Henriques2019}%
  \BibitemOpen
  \bibfield  {author} {\bibinfo {author} {\bibfnamefont {F.}~\bibnamefont
  {Henriques}}, \bibinfo {author} {\bibfnamefont {F.}~\bibnamefont {Valenti}},
  \bibinfo {author} {\bibfnamefont {T.}~\bibnamefont {Charpentier}}, \bibinfo
  {author} {\bibfnamefont {M.}~\bibnamefont {Lagoin}}, \bibinfo {author}
  {\bibfnamefont {C.}~\bibnamefont {Gouriou}}, \bibinfo {author} {\bibfnamefont
  {M.}~\bibnamefont {Mart{\'\i}nez}}, \bibinfo {author} {\bibfnamefont
  {L.}~\bibnamefont {Cardani}}, \bibinfo {author} {\bibfnamefont
  {L.}~\bibnamefont {Gr{\"u}nhaupt}}, \bibinfo {author} {\bibfnamefont
  {D.}~\bibnamefont {Gusenkova}}, \bibinfo {author} {\bibfnamefont
  {J.}~\bibnamefont {Ferrero}},  \emph {et~al.},\ }\href
  {https://arxiv.org/abs/1908.04257} {\bibfield  {journal} {\bibinfo  {journal}
  {arXiv preprint arXiv:1908.04257}\ } (\bibinfo {year} {2019})}\BibitemShut
  {NoStop}%
\bibitem [{\citenamefont {Barends}\ \emph {et~al.}(2011)\citenamefont
  {Barends}, \citenamefont {Wenner}, \citenamefont {Lenander}, \citenamefont
  {Chen}, \citenamefont {Bialczak}, \citenamefont {Kelly}, \citenamefont
  {Lucero}, \citenamefont {O'Malley}, \citenamefont {Mariantoni}, \citenamefont
  {Sank}, \citenamefont {Wang}, \citenamefont {White}, \citenamefont {Yin},
  \citenamefont {Zhao}, \citenamefont {Cleland}, \citenamefont {Martinis},\
  and\ \citenamefont {Baselmans}}]{Barends2011}%
  \BibitemOpen
  \bibfield  {author} {\bibinfo {author} {\bibfnamefont {R.}~\bibnamefont
  {Barends}}, \bibinfo {author} {\bibfnamefont {J.}~\bibnamefont {Wenner}},
  \bibinfo {author} {\bibfnamefont {M.}~\bibnamefont {Lenander}}, \bibinfo
  {author} {\bibfnamefont {Y.}~\bibnamefont {Chen}}, \bibinfo {author}
  {\bibfnamefont {R.~C.}\ \bibnamefont {Bialczak}}, \bibinfo {author}
  {\bibfnamefont {J.}~\bibnamefont {Kelly}}, \bibinfo {author} {\bibfnamefont
  {E.}~\bibnamefont {Lucero}}, \bibinfo {author} {\bibfnamefont
  {P.}~\bibnamefont {O'Malley}}, \bibinfo {author} {\bibfnamefont
  {M.}~\bibnamefont {Mariantoni}}, \bibinfo {author} {\bibfnamefont
  {D.}~\bibnamefont {Sank}}, \bibinfo {author} {\bibfnamefont {H.}~\bibnamefont
  {Wang}}, \bibinfo {author} {\bibfnamefont {T.~C.}\ \bibnamefont {White}},
  \bibinfo {author} {\bibfnamefont {Y.}~\bibnamefont {Yin}}, \bibinfo {author}
  {\bibfnamefont {J.}~\bibnamefont {Zhao}}, \bibinfo {author} {\bibfnamefont
  {A.~N.}\ \bibnamefont {Cleland}}, \bibinfo {author} {\bibfnamefont {J.~M.}\
  \bibnamefont {Martinis}}, \ and\ \bibinfo {author} {\bibfnamefont {J.~J.~A.}\
  \bibnamefont {Baselmans}},\ }\href {\doibase 10.1063/1.3638063} {\bibfield
  {journal} {\bibinfo  {journal} {Appl. Phys. Lett.}\ }\textbf {\bibinfo
  {volume} {99}},\ \bibinfo {eid} {113507} (\bibinfo {year}
  {2011})}\BibitemShut {NoStop}%
\bibitem [{\citenamefont {Yeh}\ \emph {et~al.}(2017)\citenamefont {Yeh},
  \citenamefont {LeFebvre}, \citenamefont {Premaratne}, \citenamefont
  {Wellstood},\ and\ \citenamefont {Palmer}}]{Yeh2017}%
  \BibitemOpen
  \bibfield  {author} {\bibinfo {author} {\bibfnamefont {J.-H.}\ \bibnamefont
  {Yeh}}, \bibinfo {author} {\bibfnamefont {J.}~\bibnamefont {LeFebvre}},
  \bibinfo {author} {\bibfnamefont {S.}~\bibnamefont {Premaratne}}, \bibinfo
  {author} {\bibfnamefont {F.}~\bibnamefont {Wellstood}}, \ and\ \bibinfo
  {author} {\bibfnamefont {B.}~\bibnamefont {Palmer}},\ }\href {\doibase
  10.1063/1.4984894} {\bibfield  {journal} {\bibinfo  {journal} {Journal of
  Applied Physics}\ }\textbf {\bibinfo {volume} {121}},\ \bibinfo {pages}
  {224501} (\bibinfo {year} {2017})}\BibitemShut {NoStop}%
\bibitem [{\citenamefont {Krinner}\ \emph {et~al.}(2019)\citenamefont
  {Krinner}, \citenamefont {Storz}, \citenamefont {Kurpiers}, \citenamefont
  {Magnard}, \citenamefont {Heinsoo}, \citenamefont {Keller}, \citenamefont
  {L{\"u}tolf}, \citenamefont {Eichler},\ and\ \citenamefont
  {Wallraff}}]{Krinner2019}%
  \BibitemOpen
  \bibfield  {author} {\bibinfo {author} {\bibfnamefont {S.}~\bibnamefont
  {Krinner}}, \bibinfo {author} {\bibfnamefont {S.}~\bibnamefont {Storz}},
  \bibinfo {author} {\bibfnamefont {P.}~\bibnamefont {Kurpiers}}, \bibinfo
  {author} {\bibfnamefont {P.}~\bibnamefont {Magnard}}, \bibinfo {author}
  {\bibfnamefont {J.}~\bibnamefont {Heinsoo}}, \bibinfo {author} {\bibfnamefont
  {R.}~\bibnamefont {Keller}}, \bibinfo {author} {\bibfnamefont
  {J.}~\bibnamefont {L{\"u}tolf}}, \bibinfo {author} {\bibfnamefont
  {C.}~\bibnamefont {Eichler}}, \ and\ \bibinfo {author} {\bibfnamefont
  {A.}~\bibnamefont {Wallraff}},\ }\href {\doibase
  10.1140/epjqt/s40507-019-0072-0} {\bibfield  {journal} {\bibinfo  {journal}
  {EPJ Quantum Technology}\ }\textbf {\bibinfo {volume} {6}},\ \bibinfo {pages}
  {2} (\bibinfo {year} {2019})}\BibitemShut {NoStop}%
\bibitem [{\citenamefont {Jin}\ \emph {et~al.}(2015)\citenamefont {Jin},
  \citenamefont {Kamal}, \citenamefont {Sears}, \citenamefont {Gudmundsen},
  \citenamefont {Hover}, \citenamefont {Miloshi}, \citenamefont {Slattery},
  \citenamefont {Yan}, \citenamefont {Yoder}, \citenamefont {Orlando},
  \citenamefont {Gustavsson},\ and\ \citenamefont
  {Oliver}}]{Oliver2015Residual}%
  \BibitemOpen
  \bibfield  {author} {\bibinfo {author} {\bibfnamefont {X.~Y.}\ \bibnamefont
  {Jin}}, \bibinfo {author} {\bibfnamefont {A.}~\bibnamefont {Kamal}}, \bibinfo
  {author} {\bibfnamefont {A.~P.}\ \bibnamefont {Sears}}, \bibinfo {author}
  {\bibfnamefont {T.}~\bibnamefont {Gudmundsen}}, \bibinfo {author}
  {\bibfnamefont {D.}~\bibnamefont {Hover}}, \bibinfo {author} {\bibfnamefont
  {J.}~\bibnamefont {Miloshi}}, \bibinfo {author} {\bibfnamefont
  {R.}~\bibnamefont {Slattery}}, \bibinfo {author} {\bibfnamefont
  {F.}~\bibnamefont {Yan}}, \bibinfo {author} {\bibfnamefont {J.}~\bibnamefont
  {Yoder}}, \bibinfo {author} {\bibfnamefont {T.~P.}\ \bibnamefont {Orlando}},
  \bibinfo {author} {\bibfnamefont {S.}~\bibnamefont {Gustavsson}}, \ and\
  \bibinfo {author} {\bibfnamefont {W.~D.}\ \bibnamefont {Oliver}},\ }\href
  {\doibase 10.1103/PhysRevLett.114.240501} {\bibfield  {journal} {\bibinfo
  {journal} {Phys. Rev. Lett.}\ }\textbf {\bibinfo {volume} {114}},\ \bibinfo
  {pages} {240501} (\bibinfo {year} {2015})}\BibitemShut {NoStop}%
\bibitem [{\citenamefont {Liu}(2014)}]{Liu2014}%
  \BibitemOpen
  \bibfield  {author} {\bibinfo {author} {\bibfnamefont {Y.}~\bibnamefont
  {Liu}},\ }\href {\doibase 10.22661/AAPPSBL.2014.24.3.05} {\bibfield
  {journal} {\bibinfo  {journal} {AAPPS Bull.}\ }\textbf {\bibinfo {volume}
  {24}},\ \bibinfo {pages} {5} (\bibinfo {year} {2014})}\BibitemShut {NoStop}%
\bibitem [{\citenamefont {Blais}\ \emph {et~al.}(2004)\citenamefont {Blais},
  \citenamefont {Huang}, \citenamefont {Wallraff}, \citenamefont {Girvin},\
  and\ \citenamefont {Schoelkopf}}]{Blais2004}%
  \BibitemOpen
  \bibfield  {author} {\bibinfo {author} {\bibfnamefont {A.}~\bibnamefont
  {Blais}}, \bibinfo {author} {\bibfnamefont {R.-S.}\ \bibnamefont {Huang}},
  \bibinfo {author} {\bibfnamefont {A.}~\bibnamefont {Wallraff}}, \bibinfo
  {author} {\bibfnamefont {S.~M.}\ \bibnamefont {Girvin}}, \ and\ \bibinfo
  {author} {\bibfnamefont {R.~J.}\ \bibnamefont {Schoelkopf}},\ }\href
  {\doibase 10.1103/PhysRevA.69.062320} {\bibfield  {journal} {\bibinfo
  {journal} {Phys. Rev. A}\ }\textbf {\bibinfo {volume} {69}},\ \bibinfo
  {pages} {062320} (\bibinfo {year} {2004})}\BibitemShut {NoStop}%
\bibitem [{\citenamefont {Eichler}\ \emph {et~al.}(2014)\citenamefont
  {Eichler}, \citenamefont {Salathe}, \citenamefont {Mlynek}, \citenamefont
  {Schmidt},\ and\ \citenamefont {Wallraff}}]{Eichler2014}%
  \BibitemOpen
  \bibfield  {author} {\bibinfo {author} {\bibfnamefont {C.}~\bibnamefont
  {Eichler}}, \bibinfo {author} {\bibfnamefont {Y.}~\bibnamefont {Salathe}},
  \bibinfo {author} {\bibfnamefont {J.}~\bibnamefont {Mlynek}}, \bibinfo
  {author} {\bibfnamefont {S.}~\bibnamefont {Schmidt}}, \ and\ \bibinfo
  {author} {\bibfnamefont {A.}~\bibnamefont {Wallraff}},\ }\href {\doibase
  http://dx.doi.org/10.1103/PhysRevLett.113.110502} {\bibfield  {journal}
  {\bibinfo  {journal} {Phys. Rev. Lett.}\ ,\ \bibinfo {pages} {110502}}
  (\bibinfo {year} {2014})}\BibitemShut {NoStop}%
\bibitem [{Sup()}]{Supp}%
  \BibitemOpen
  \href@noop {} {}\bibinfo {note} {See Supplemental Material at [URL will be
  inserted by publisher] for the derivations of the expressions used,
  experimental details, additional data and detailed discussion of errors. It
  includes Refs. [??-??].}\BibitemShut {Stop}%
\bibitem [{\citenamefont {Aumentado}\ \emph {et~al.}(2004)\citenamefont
  {Aumentado}, \citenamefont {Keller}, \citenamefont {Martinis},\ and\
  \citenamefont {Devoret}}]{Aumentado2004}%
  \BibitemOpen
  \bibfield  {author} {\bibinfo {author} {\bibfnamefont {J.}~\bibnamefont
  {Aumentado}}, \bibinfo {author} {\bibfnamefont {M.~W.}\ \bibnamefont
  {Keller}}, \bibinfo {author} {\bibfnamefont {J.~M.}\ \bibnamefont
  {Martinis}}, \ and\ \bibinfo {author} {\bibfnamefont {M.~H.}\ \bibnamefont
  {Devoret}},\ }\href@noop {} {\bibfield  {journal} {\bibinfo  {journal} {Phys.
  Rev. Lett.}\ }\textbf {\bibinfo {volume} {92}},\ \bibinfo {pages} {066802}
  (\bibinfo {year} {2004})}\BibitemShut {NoStop}%
\bibitem [{\citenamefont {Catelani}\ \emph {et~al.}(2011)\citenamefont
  {Catelani}, \citenamefont {Koch}, \citenamefont {Frunzio}, \citenamefont
  {Schoelkopf}, \citenamefont {Devoret},\ and\ \citenamefont
  {Glazman}}]{Catelani2011}%
  \BibitemOpen
  \bibfield  {author} {\bibinfo {author} {\bibfnamefont {G.}~\bibnamefont
  {Catelani}}, \bibinfo {author} {\bibfnamefont {J.}~\bibnamefont {Koch}},
  \bibinfo {author} {\bibfnamefont {L.}~\bibnamefont {Frunzio}}, \bibinfo
  {author} {\bibfnamefont {R.~J.}\ \bibnamefont {Schoelkopf}}, \bibinfo
  {author} {\bibfnamefont {M.~H.}\ \bibnamefont {Devoret}}, \ and\ \bibinfo
  {author} {\bibfnamefont {L.~I.}\ \bibnamefont {Glazman}},\ }\href {\doibase
  10.1103/PhysRevLett.106.077002} {\bibfield  {journal} {\bibinfo  {journal}
  {Phys. Rev. Lett.}\ }\textbf {\bibinfo {volume} {106}},\ \bibinfo {pages}
  {077002} (\bibinfo {year} {2011})}\BibitemShut {NoStop}%
\bibitem [{\citenamefont {Klimov}\ \emph {et~al.}(2018)\citenamefont {Klimov},
  \citenamefont {Kelly}, \citenamefont {Chen}, \citenamefont {Neeley},
  \citenamefont {Megrant}, \citenamefont {Burkett}, \citenamefont {Barends},
  \citenamefont {Arya}, \citenamefont {Chiaro}, \citenamefont {Chen},
  \citenamefont {Dunsworth}, \citenamefont {Fowler}, \citenamefont {Foxen},
  \citenamefont {Gidney}, \citenamefont {Giustina}, \citenamefont {Graff},
  \citenamefont {Huang}, \citenamefont {Jeffrey}, \citenamefont {Lucero},
  \citenamefont {Mutus}, \citenamefont {Naaman}, \citenamefont {Neill},
  \citenamefont {Quintana}, \citenamefont {Roushan}, \citenamefont {Sank},
  \citenamefont {Vainsencher}, \citenamefont {Wenner}, \citenamefont {White},
  \citenamefont {Boixo}, \citenamefont {Babbush}, \citenamefont {Smelyanskiy},
  \citenamefont {Neven},\ and\ \citenamefont {Martinis}}]{Klimov2018}%
  \BibitemOpen
  \bibfield  {author} {\bibinfo {author} {\bibfnamefont {P.~V.}\ \bibnamefont
  {Klimov}}, \bibinfo {author} {\bibfnamefont {J.}~\bibnamefont {Kelly}},
  \bibinfo {author} {\bibfnamefont {Z.}~\bibnamefont {Chen}}, \bibinfo {author}
  {\bibfnamefont {M.}~\bibnamefont {Neeley}}, \bibinfo {author} {\bibfnamefont
  {A.}~\bibnamefont {Megrant}}, \bibinfo {author} {\bibfnamefont
  {B.}~\bibnamefont {Burkett}}, \bibinfo {author} {\bibfnamefont
  {R.}~\bibnamefont {Barends}}, \bibinfo {author} {\bibfnamefont
  {K.}~\bibnamefont {Arya}}, \bibinfo {author} {\bibfnamefont {B.}~\bibnamefont
  {Chiaro}}, \bibinfo {author} {\bibfnamefont {Y.}~\bibnamefont {Chen}},
  \bibinfo {author} {\bibfnamefont {A.}~\bibnamefont {Dunsworth}}, \bibinfo
  {author} {\bibfnamefont {A.}~\bibnamefont {Fowler}}, \bibinfo {author}
  {\bibfnamefont {B.}~\bibnamefont {Foxen}}, \bibinfo {author} {\bibfnamefont
  {C.}~\bibnamefont {Gidney}}, \bibinfo {author} {\bibfnamefont
  {M.}~\bibnamefont {Giustina}}, \bibinfo {author} {\bibfnamefont
  {R.}~\bibnamefont {Graff}}, \bibinfo {author} {\bibfnamefont
  {T.}~\bibnamefont {Huang}}, \bibinfo {author} {\bibfnamefont
  {E.}~\bibnamefont {Jeffrey}}, \bibinfo {author} {\bibfnamefont
  {E.}~\bibnamefont {Lucero}}, \bibinfo {author} {\bibfnamefont {J.~Y.}\
  \bibnamefont {Mutus}}, \bibinfo {author} {\bibfnamefont {O.}~\bibnamefont
  {Naaman}}, \bibinfo {author} {\bibfnamefont {C.}~\bibnamefont {Neill}},
  \bibinfo {author} {\bibfnamefont {C.}~\bibnamefont {Quintana}}, \bibinfo
  {author} {\bibfnamefont {P.}~\bibnamefont {Roushan}}, \bibinfo {author}
  {\bibfnamefont {D.}~\bibnamefont {Sank}}, \bibinfo {author} {\bibfnamefont
  {A.}~\bibnamefont {Vainsencher}}, \bibinfo {author} {\bibfnamefont
  {J.}~\bibnamefont {Wenner}}, \bibinfo {author} {\bibfnamefont {T.~C.}\
  \bibnamefont {White}}, \bibinfo {author} {\bibfnamefont {S.}~\bibnamefont
  {Boixo}}, \bibinfo {author} {\bibfnamefont {R.}~\bibnamefont {Babbush}},
  \bibinfo {author} {\bibfnamefont {V.~N.}\ \bibnamefont {Smelyanskiy}},
  \bibinfo {author} {\bibfnamefont {H.}~\bibnamefont {Neven}}, \ and\ \bibinfo
  {author} {\bibfnamefont {J.~M.}\ \bibnamefont {Martinis}},\ }\href {\doibase
  10.1103/PhysRevLett.121.090502} {\bibfield  {journal} {\bibinfo  {journal}
  {Phys. Rev. Lett.}\ }\textbf {\bibinfo {volume} {121}},\ \bibinfo {pages}
  {090502} (\bibinfo {year} {2018})}\BibitemShut {NoStop}%
\bibitem [{\citenamefont {Motzoi}\ \emph {et~al.}(2009)\citenamefont {Motzoi},
  \citenamefont {Gambetta}, \citenamefont {Rebentrost},\ and\ \citenamefont
  {Wilhelm}}]{Motzoi2009}%
  \BibitemOpen
  \bibfield  {author} {\bibinfo {author} {\bibfnamefont {F.}~\bibnamefont
  {Motzoi}}, \bibinfo {author} {\bibfnamefont {J.~M.}\ \bibnamefont
  {Gambetta}}, \bibinfo {author} {\bibfnamefont {P.}~\bibnamefont
  {Rebentrost}}, \ and\ \bibinfo {author} {\bibfnamefont {F.~K.}\ \bibnamefont
  {Wilhelm}},\ }\href {\doibase 10.1103/PhysRevLett.103.110501} {\bibfield
  {journal} {\bibinfo  {journal} {Phys. Rev. Lett.}\ }\textbf {\bibinfo
  {volume} {103}},\ \bibinfo {eid} {110501} (\bibinfo {year}
  {2009})}\BibitemShut {NoStop}%
\bibitem [{not()}]{note_snr}%
  \BibitemOpen
  \href@noop {} {}\bibinfo {note} {The precision $\sigma$ for a correlator
  measurement scales as $\sigma \propto N^{1/2}$. But also being a correlator
  entails $\sigma \propto r^{2}$, where $r$ is a signal-to-noise ratio for a
  single read out. Therefore having the same precision $\sigma = const$ leads
  to $N^{1/2}r^2 = const$, or $N \propto r^{-4}$}\BibitemShut {NoStop}%
\bibitem [{\citenamefont {Chen}\ \emph {et~al.}(2016)\citenamefont {Chen},
  \citenamefont {Kelly}, \citenamefont {Quintana}, \citenamefont {Barends},
  \citenamefont {Campbell}, \citenamefont {Chen}, \citenamefont {Chiaro},
  \citenamefont {Dunsworth}, \citenamefont {Fowler}, \citenamefont {Lucero},
  \citenamefont {Jeffrey}, \citenamefont {Megrant}, \citenamefont {Mutus},
  \citenamefont {Neeley}, \citenamefont {Neill}, \citenamefont {O'Malley},
  \citenamefont {Roushan}, \citenamefont {Sank}, \citenamefont {Vainsencher},
  \citenamefont {Wenner}, \citenamefont {White}, \citenamefont {Korotkov},\
  and\ \citenamefont {Martinis}}]{Martinis2016leakage}%
  \BibitemOpen
  \bibfield  {author} {\bibinfo {author} {\bibfnamefont {Z.}~\bibnamefont
  {Chen}}, \bibinfo {author} {\bibfnamefont {J.}~\bibnamefont {Kelly}},
  \bibinfo {author} {\bibfnamefont {C.}~\bibnamefont {Quintana}}, \bibinfo
  {author} {\bibfnamefont {R.}~\bibnamefont {Barends}}, \bibinfo {author}
  {\bibfnamefont {B.}~\bibnamefont {Campbell}}, \bibinfo {author}
  {\bibfnamefont {Y.}~\bibnamefont {Chen}}, \bibinfo {author} {\bibfnamefont
  {B.}~\bibnamefont {Chiaro}}, \bibinfo {author} {\bibfnamefont
  {A.}~\bibnamefont {Dunsworth}}, \bibinfo {author} {\bibfnamefont {A.~G.}\
  \bibnamefont {Fowler}}, \bibinfo {author} {\bibfnamefont {E.}~\bibnamefont
  {Lucero}}, \bibinfo {author} {\bibfnamefont {E.}~\bibnamefont {Jeffrey}},
  \bibinfo {author} {\bibfnamefont {A.}~\bibnamefont {Megrant}}, \bibinfo
  {author} {\bibfnamefont {J.}~\bibnamefont {Mutus}}, \bibinfo {author}
  {\bibfnamefont {M.}~\bibnamefont {Neeley}}, \bibinfo {author} {\bibfnamefont
  {C.}~\bibnamefont {Neill}}, \bibinfo {author} {\bibfnamefont {P.~J.~J.}\
  \bibnamefont {O'Malley}}, \bibinfo {author} {\bibfnamefont {P.}~\bibnamefont
  {Roushan}}, \bibinfo {author} {\bibfnamefont {D.}~\bibnamefont {Sank}},
  \bibinfo {author} {\bibfnamefont {A.}~\bibnamefont {Vainsencher}}, \bibinfo
  {author} {\bibfnamefont {J.}~\bibnamefont {Wenner}}, \bibinfo {author}
  {\bibfnamefont {T.~C.}\ \bibnamefont {White}}, \bibinfo {author}
  {\bibfnamefont {A.~N.}\ \bibnamefont {Korotkov}}, \ and\ \bibinfo {author}
  {\bibfnamefont {J.~M.}\ \bibnamefont {Martinis}},\ }\href {\doibase
  10.1103/PhysRevLett.116.020501} {\bibfield  {journal} {\bibinfo  {journal}
  {Phys. Rev. Lett.}\ }\textbf {\bibinfo {volume} {116}},\ \bibinfo {pages}
  {020501} (\bibinfo {year} {2016})}\BibitemShut {NoStop}%
\end{thebibliography}%
%\bibliography{R:/EQUS-SQDLab/RefDB/SQDRefDB}

\clearpage

\subsection{Supplementary material}

%{\it Theoretical idea.---}
Equation~\eqref{eqn:p_e_base} can be derived from the following system of equations:
\begin{equation*}
\left\{\begin{aligned}
&P_g\cdot V_g^2 + P_e\cdot V_e^2 = g^{(1)}(0)\\
&P_g^2\cdot V_g^2 + 2P_gP_eV_gV_e + P_e^2\cdot V_e^2 = g^{(1)}(\infty) \\
&\frac{1}{4}(V_g^2 + 2V_gV_g + V_e^2) = \frac{1}{4}\left(g^{(0)}+g^{(0)}_\pi\right)^2 \equiv g^{(1)}_{\pi/2}(\infty) \\
&P_g + P_e = 1
\end{aligned} \right.
\end{equation*}

Assuming that $P_e$ is small we write the system to the lowest order in $P_e$ as:
\begin{equation}
\left\{\begin{aligned}
&V_g^2 + P_e (V_e^2 - V_g^2) = g^{(1)}(0)\\
&V_g^2 + 2 P_e (V_gV_e - V_g^2) = g^{(1)}(\infty) \\
&(V_g + V_e)^2 = 4g^{(1)}_{\pi/2}(\infty)
\end{aligned} \right.
%\quad \text{Resulting eqs}
\label{Supp:eqns}
\end{equation}

The difference of two correlators reads
\eqn{\begin{aligned} &g^{(1)}(\infty) - g^{(1)}(0) = \\ & P_e \left[2V_g(V_e-V_g) - (V_e - V_g)(V_e + V_g)\right] = \\ &=   P_e (V_e-V_g)(V_g-V_e) \simeq \\ &\sim P_e\left(2\sqrt{g^{(1)}_{\pi/2}(\infty)} - 2\sqrt{g^{(1)}(0)}\right)^2, \end{aligned}}
where we utilized the smallness of $P_e$ to write $V_g \sim \sqrt{g^{(1)}(0)}$ and used the third equation from \eqref{Supp:eqns}. It follows that
\eqnl{eqn:p_e_base:supp}{P_e \simeq \frac{g^{(1)}(0)-g^{(1)}(\infty)}{4\left(\sqrt{g^{(1)}_{\pi/2}(\infty)} - \sqrt{g^{(1)}(0)}\right)^2}.}

Note, that only measuring $g^{(1)}(0)$ requires actual correlation of two sequential readouts. Both $g^{(1)}(\infty)$ and $g^{(1)}_{\pi/2}(\infty)$ can be obtained by squaring the mean values of the ground state and 50/50 mixture of $\g$ and $\e$ states, respectively.

{\it Experimental details.---} The number of averages $N$ required to reach a certain final precision for a first order correlation function  scales as $N^{-4}$. This scaling is unfavorable  compared to $N^{-2}$ for a zeroth order correlation function~\cite{note_snr} and requires additional measures to optimize the SNR.

We first used a quantum limited JPA to amplify the signal to a high SNR of $\approx 6$ in order to classify the measurement outcomes in a single shot. After performing single shot measurements and state classification, the data is converted to the binary form ($\g \ra 0;\e \ra 1$). The procedure allows for measurement of decay of the correlation function $g^{(1)}(\tau)$ and extraction of the excited state population without the noise of the amplification chain. As expected, the result coincided perfectly with the direct counts of $\g$ and $\e$ states.

\begin{figure}[t]
	\centering
	\includegraphics[width=0.49\textwidth]{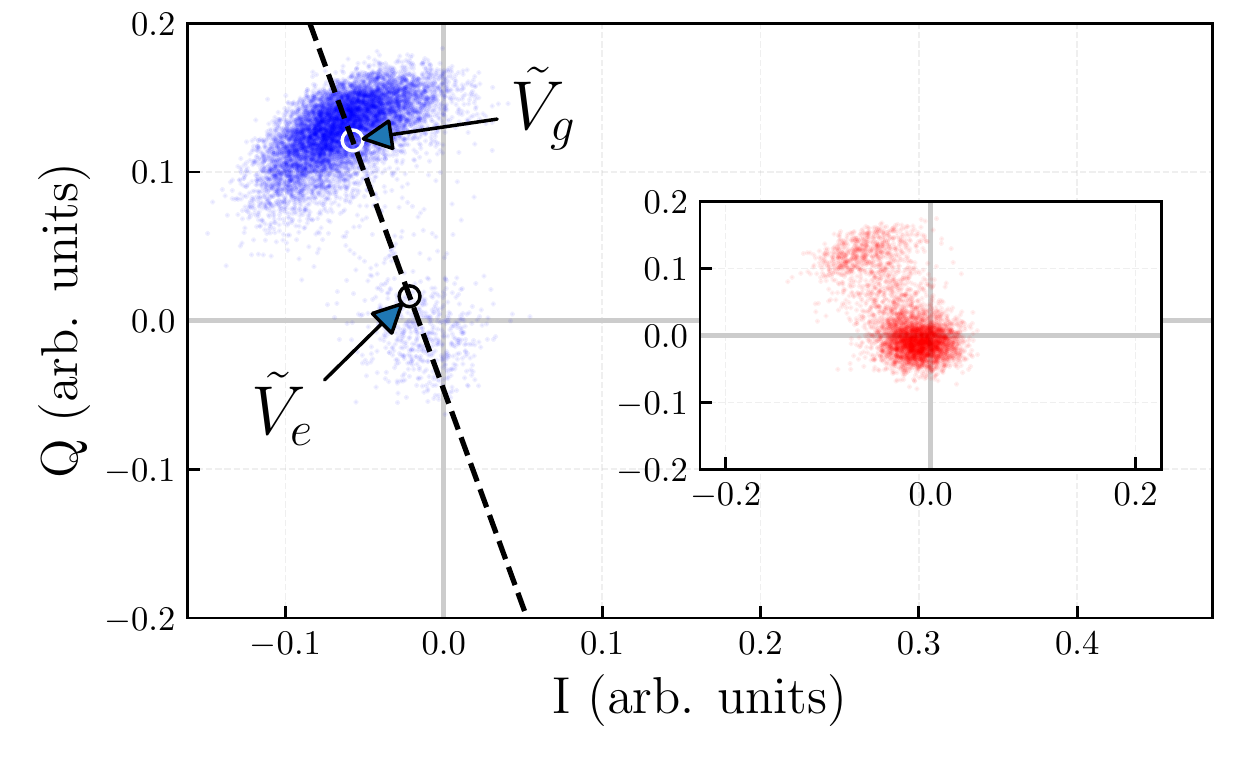}
	\caption{Single-shot measurement data with no pulse ($\pi$-pulse) shown as blue (red) pixels. The white (black) circle shows the averaged response $\tilde V_g$ ($\tilde V_e$). Dashed is the line joining $V_g$ and $V_e$; We assume deviations perpendicular to this line are only caused by noise, and project the complex IQ measurement onto it.}
	\label{fig:snr_IQ}
\end{figure}

In order to make the method work without the state classification, we use the integrated heterodyne voltage $V$ to extract a real-valued response containing a maximum amount of information. First, we do a standard set of calibration measurements to determine $\tilde V_g$ and $\tilde V_e$, which are complex values of integrated heterodyne voltage. Each of the subsequent measurements we project onto the line containing the two calibration responses, such that $\tilde V_g \ra 0$ and $\tilde V_e \ra 1$  (see Fig.~\ref{fig:snr_IQ}). This is done merely by the transformation mentioned in the main text:
\eqn{{\bar V} = \operatorname{Re} \left[ (V-\tilde V_g)/(\tilde V_e-\tilde V_g) \right].
\label{eqn:transformation:supp}}

%\eqnl{eqn:p_e_base:tmp}{P_e \approx \frac{\bar{g}^{(1)}(0) - \bar{g}^{(1)}(\infty)}{4\left(\sqrt{\bar{g}^{(1)}_{\pi/2}(\infty)} - \sqrt{\bar{g}^{(1)}(0)}\right)^2}.}

In this case Eq.~(\ref{eqn:p_e_base:supp}) takes a particularly simple form 
\eqnl{eqn:p_e_intermediate}{P_e \simeq \frac{\bar{g}^{(1)}(0)}{4\left(\sqrt{1/4} - \sqrt{\bar{g}^{(1)}(0)}\right)^2},}
where we used $\bar{g}^{(1)}(\infty) = 0$ and $\bar{g}^{(1)}_{\pi/2}(\infty) = 1/4$. Moreover, after the transformation $\bar{g}^{(1)}(0)$ is close to zero since the qubit is mostly in the ground state. Any error in the estimation of $\bar{g}^{(1)}(0)$ will be amplified due to the square root operation of a smal number in the expression for excited state population (Eq.~(\ref{eqn:p_e_intermediate})). Removing $\sqrt{\bar{g}^{(1)}(0)}$ in the denominator will not lead to inaccuracy for a small excited state population but will greatly reduce noise in $\bar{g}^{(1)}(0)$. 
%Another reason for the removal of the $\sqrt{\bar{g}^{(1)}(0)}$ term is explained below, and has to do with slow noise in the equipment. 
These manipulations lead to the following expression for $P_e$:
\eqn{P_e\simeq\bar{g}^{(1)}(0).}

Alternatively, one can re-derive the expression for $P_e$ using the normalised responses and obtain an exact solution for the system of equations with three unknowns:
\begin{equation*}
\left\{\begin{aligned}
&P_g\cdot \bar{V}_g^2 + P_e\cdot \bar{V}_e^2 = \bar{g}^{(1)}(0)\\
&P_g^2\cdot \bar{V}_g^2 + 2P_gP_e\bar{V}_g\bar{V}_e + P_e^2\cdot \bar{V}_e^2 = 0 \\
&\frac{1}{4}(\bar{V}_g^2 + 2\bar{V}_g\bar{V}_e + \bar{V}_e^2) = 0.5 \\
&P_g + P_e = 1
\end{aligned} \right.
\end{equation*}
Note that after the normalization, $\tilde V_g \ra 0$ and $\tilde V_e \ra 1$ but the 'true' responses $V_g \ra \bar{V}_g$ and $V_e \ra \bar{V}_e$ are still unknown.
Solving the system leads to a precise expression:
\eqn{P_e = \frac{1}{2}-\frac{1}{2 \sqrt{1+4 \bar{g}^{(1)}(0)}}.}

\subsection{Precision of the conventional methods}

%In this section we briefly compare the correlator method to the two widely used ones, in particular their sensitivity to various types of error sources.

%{\it Precision and errors.---}
In this section we present definition of systematic and statistical errors and analyze them in more details for single-shot counting and for the method exploiting the third level (qutrit).

We are going to distinguish two sorts of errors with respect to precision:

-- Relative errors, i.e. errors causing under- or overestimation of the result by some fraction. These errors may limit the relative precision of a method, but they do not pose an absolute precision floor.

-- Absolute errors, i.e. errors which do not depend on the $\e$-state population of a qubit, but rather on other setup parameters, such as the system's SNR, measurement time or setup stability.

While of course the magnitude of the errors under consideration is the primary factor, we generally note that the absolute errors are more detrimental as they limit both relative and absolute precision and eventually lead to a limit on the lowest measurable T$_{\rm eff}$. Relative errors, on the other hand, in principle allow distinguishing arbitrary low $\e$-state populations.

{\it Direct counting.---} Direct counting of the occurrences of $\e$ state in the prepared ground state of a qubit is the easiest and the most natural way to obtain the $\e$-state population, and effective temperature, of the qubit. With high signal-to-noise ratios of 10+ it offers the direct and the most reliable way to do so while also allowing one to immediately see and quantify all error sources.

Its statistical error scales as the one for a single measurement and therefore decays quickly with the amount of repetitions. A figure here would be the standard deviation of the amount of $\e$ counts in the prepared ground state, and therefore it naturally does not have a noise floor. There are three main causes of systematic errors, however: normal $T_1$-decay (or excitation) during the readout, measurement backaction and misinterpretation of the outcomes falling outside of the state line boundary.

The $T_1$-decay possesses two important properties allowing us to integrate it into the model and disregard entirely. First, it is a relative error and can be estimated precisely both in case of decay and excitation. And second, in case of equilibrium $\e$-state population the rate of decays equals the rate of excitations, and therefore this error does not play any role in the total error budget.

Misinterpretation is another straightforward error to consider qualitatively. It is obvious that this error is absolute, as it does not depend in the first order on the $\e$-state population, but depends only on the SNR. As an example, if the SNR is 6, the decision boundary between g-state and $\e$-state would be at the distance of three sigmas from the "true" values of both $\g$ and $\e$, which would lead to an amount on the order of "three sigma error" -- that is, 0.3\% in this case -- to be misinterpreted. Of course, it is possible to employ various techniques to mitigate those errors somehow, but the key property -- the errors being absolute -- stays. Non-Gaussian noise distribution due to multiplied noise from the amplification chain and room-temperature elements presents an extra complication in realization of the mentioned mitigation techniques. Extra assumptions on the origins and distribution of the readout noise should be used, which can prove to be a challenging task.% [some thesis].

To sum up, the statistical error following this way is sufficiently small after only a few repetitions, and generally one just requires an ample amount of $\e$ counts to infer the population. From this we can say that statistical error scaling in this way is the best out of the three presented methods. However, the systematic errors due to misinterpretation of the outcomes pose an eventual absolute precision floor on the accuracy of the method.

{\it Qutrit method.---} In addition to the requirement of going out of the qubit subspace and exploiting the higher ($\f$) level of a physical system, the qutrit method suffers from a few extra error channels which are largely absent in both the correlator and single-shot method cases.

The small errors $\varepsilon$ of calibration pulses, such as $\pi$-pulse in the correlator method or the $pi$-pulse between $\g$ and $\e$ levels in the qutrit method, contribute to the total error budget only in the second order and create a relative error. It is straightforward to see, as those errors go to the denominator of the corresponding expressions to determine the $\e$-state population, thus multiplying the scale of the measurement by a factor $1 + \varepsilon$ close to unity.
An $\e$-$\f$ $\pi$-pulse in the qutrit method, however, creates a small first-order error by directly exciting the $\e$ state of the qutrit due to the coupling of the drive to the $\g$-$\e$ transition. Another reason is the envelope of the $\e$-$\f$ pulse in the frequency domain, which has finite bandwidth commensurate with the anharmonicity of the qubit (usually on the order of tens per cent). This unwanted excitation of $\e$ state during the measurement of its population is an example of a leakage error \cite{Martinis2016leakage} present in the qutrit protocol.

%\cite{Martinis2016leakage} -- paper by Martinis about leakage

Therefore calibrating the e-f pulse so that in the Fourier space no g-e frequency component is present becomes imperative for the qutrit method in order to measure low populations. While certain optimal control techniques, such as DRAG pulses \cite{Motzoi2009} for e-f transition, could be employed to mitigate this problem, it is not possible to entirely isolate the frequencies and therefore it poses an extra limitation on the precision of the method. Note that this error is absolute and systematic, and the decrease of statistical uncertainty can not guarantee its negligible magnitude or give information about its value.

%Fig.~\ref{fig:pop_vs_freq1} shows an increase in the effective temperature of the qubit slightly below 6 GHz, which we attribute to coupling to two-level systems. This conclusion is supported by previously observed anticrossings of the qubit with TLSs at these frequencies, yet more investigation is required to determine its source with certainty. The coupling of the $\e$-$f$ transition to two-level fluctuators leads to instability in its frequency and hybridisation of the transition, thus enhancing its decoherence and preventing the statistical error from decreasing. It might be a primary reason of the largest absolute discrepancies in measured $\e$-state population showed by the qutrit method.
\end{document}